\documentclass[useAMS,usenatbib,usegraphicx]{mn2e} 
\usepackage{amsmath,amssymb}
\usepackage{graphicx,mathptm} 
\usepackage{times} 
\usepackage{natbib}
\usepackage{psfig}

\title[Halo stochasticity in global clustering analysis]{Halo stochasticity in global clustering analysis}
\author[S. Bonoli and U.L.
Pen]{S.Bonoli,$^{1}$\thanks{E-mail:bonoli@mpa-garching.mpg.de} and
U.L.Pen,$^{2}$\thanks{E-mail:pen@cita.utoronto.ca} \\
$^{1}$ Max-Planck-Institut f\"ur Astrophysik, Karl-Schwarzschild-Str. 1, D85748
Garching, Germany\\
$^{2}$Canadian Institute for Theoretical Astrophysics, 60 St. George Street, Toronto, M5S 3H8, Canada}
\begin{document}

\date{}

\pagerange{\pageref{firstpage}--\pageref{lastpage}} \pubyear{2006}

\maketitle

\label{firstpage}

\begin{abstract}  
  Galaxy clustering and cosmic magnification can be used to estimate the dark
  matter power spectrum if the theoretical relation between the distribution of
  galaxies and the distribution of dark matter is precisely known. In the
  present work we study the statistics of haloes, which in the halo model
  determines the distribution of galaxies.  Haloes are known to be biased
  tracer of dark matter, and at large scales it is usually assumed there
  is no intrinsic stochasticity between the two fields (i.e., $r=1$).
  Following the work of \citet[][]{seljak04}, we explore how correct this
  assumption is and, moving a step further, we try to qualify the nature of
  stochasticity.  We use Principal Component Analysis applied to the outputs of
  a cosmological N-body simulation as a function of mass to: (1) explore the
  behaviour of stochasticity in the correlation between haloes of different
  masses; (2) explore the behaviour of stochasticity in the correlation between
  haloes and dark matter.  We show results obtained using a catalogue with
  $2.1$ million haloes, from a PMFAST simulation with box size of
  $1000h^{-1}$Mpc and with about $4$ billion particles.  

In the relation between different populations of haloes we find that stochasticity is
  not-negligible even at large scales. In agreement with the conclusions of
\citet{tegmark99} who studied the correlations of different galaxy populations,
we found that the shot-noise subtracted stochasticity is
  qualitatively different from `enhanced' shot noise and, specifically, it is dominated by
  a single stochastic eigenvalue. We call this the `minimally stochastic'
  scenario,  as opposed to shot noise which is `maximally stochastic'. 
In the correlation between haloes and dark
  matter, we find that stochasticity is minimized, as expected, near the dark matter peak 
($k \sim 0.02 ~ h ~ \rm{Mpc}^{-1}$ for a $\Lambda$CDM cosmology), and, even at large
  scales, it is of the order of $15$ per cent above the shot noise.  Moreover, we find
  that the reconstruction of the dark matter distribution is improved when we
  use eigenvectors as tracers of the bias, but still the reconstruction is not
  perfect, due to stochasticity.
\end{abstract}


\section{Introduction}

The observational determination of the dark matter distribution is important
not only to constrain cosmological parameters, but also to understand galaxy
formation and the relation between the dark matter and the galaxy
distributions. The dark matter distribution can either be estimated indirectly
through the study of the galaxy distribution, or directly through weak
gravitational lensing.  Using the first approach, many galaxy redshift surveys,
such as 2dFGRS \citep[e.g.,][]{peacock01} and SDSS \citep[e.g.,][]{tegmark04a},
have mapped the three-dimensional distribution of around a million galaxies
to determine the real-space power spectra $P(k)$ of the matter fluctuations.  
Results from these surveys, together with the
measurements of the CMB by WMAP, favour a flat, dark-energy dominated cosmology
\citep[][]{spergel03, tegmark04b}. Of course, when using maps of galaxies to
determine the dark matter distribution, one has to take into account that
galaxies are a biased tracer of dark matter, that the
bias depends on galaxy properties and it can be scale-dependent and {\itshape
  stochastic}.  

\citet[][hereafter SW04]{seljak04} proposed to use faint galaxies as dark
matter tracer, since these are expected to occupy low mass haloes, which have a
large scale bias approximately independent of halo mass
\citep[e.g.,][]{mo96, sheth01}.  
\citet{pen04} suggested to get a dark matter three-dimensional
map and power spectrum using galaxy tomography, i.e., combining projected
weak-lensing with the cross-correlation between galaxies with distance
information (from galaxy surveys).  Galaxy-mass correlation can also be
measured using cosmic magnification, which is the magnification of background
sources due to the foreground matter distribution. \citet{scranton05} 
detected a cosmic magnification signal correlating foreground
galaxies with background quasars.  \citet{zhang05, zhang06} proposed to study
cosmic magnification using $21$ cm emitting galaxies.  All these approaches
assume a perfect correlation ($r =1 $) between galaxies and dark matter and
between galaxies of different luminosities. We stress here that with the term
\textit{stochasticity} we refer to the scatter in the correlation between two
fields (we refer to Section \ref{sec:bs} for a more rigorous definition),
without implication on the deterministic nature of the universe.  This
is analogous to the quantum mechanical density matrix.  In analogy to
the density matrix, stochasticity corresponds to a mixed state.
When there are multiple bins, for example in mass, deterministic means
a pure state, which is determined by a single state vector.  Being not
deterministic opens the dimensionality of the problem, and this paper
quantifies this effect.

On the observational side,
\citet{wild05} found stochasticity between different galaxy populations, both
when defined by colour or spectral type. On the theoretical side, SW04 found
that stochasticity is not negligible both between haloes and dark matter and
between haloes of different masses. Following SW04, we explore in more details
the relative bias and stochasticity between haloes populations, with the goal of
estimating the relative importance of this scatter. We use here a method,
based on Principal Component Analysis (PCA), to isolate the stochastic signal.
The same method was used by \citet[hereafter TB99][]{tegmark99} to study stochasticity between
different galaxy populations in the Las Campanas Redshift Survey.
This method is here tested using halo catalogues from N-body simulations
performed using the PMFAST code \citep[][]{merz05}, where we assumed that
higher-mass haloes correspond to higher-luminosity galaxies, as justified by
the tight relation between the halo mass and luminosity \citep[][]{guzik02,
hoekstra05, mandelbaum06, vandenbosch07}.  
We compare the stochastic signal with the shot noise, which is a well understood
stochastic field, and stress the fundamental
difference of our findings with a shot noise model. 

In Section 2 we give a brief review of the definitions of the
parameters used in this paper. In Section 3 we describe the simulations and the
generation of the halo catalogue. In Section 4 are presented the results of the
correlation between haloes of different mass. The bias and the stochasticity
between haloes and dark matter are described in Section 5. Finally we summarize
our conclusions in Section 6.

\section{Bias and Stochasticity}\label{sec:bs}

Observationally it has long been understood that galaxies of different masses
and colours have different clustering properties \citep[e.g., the recent works
of:][]{norberg02, zehavi05, meneux06, wang07, swanson08}.  Similarly,
it has been understood theoretically that haloes of different mass are mutually
biased.  In recent years, this has been generalized to allow for stochasticity
in addition to bias: a better quantitative model of biasing,
\textit{stochasticity},
allows a more accurate reconstruction of cosmological parameters and, even more
importantly, a better understanding of errors \citep[e.g.,][]{pen98,dekel99}.

These concepts have been primarily discussed in the context of two populations.
In this paper, we will generalize this to a continuum distribution of
populations, or bins.  The generalization of bias might appear straightforward,
but a consistent and optimal measure must be introduced.  Stochasticity is even
more complex, and potentially an open ended statistical description.

Stochasticity is a small effect, which describes a lack of coherence between 
populations.  In a Principal Component Analysis framework there are different 
possible outcomes: 1. one might find that a single parameter (eigenvector) describes 
the apparent stochasticity between all pairs of populations.  We call
this `minimally stochastic', since a single second parameters accounts
for most of the stochasticity.
Or 2. the stochasticity 
might be pluralistic, and might not be captured in one or a small number of
components.  An example of the latter is shot noise, which is pairwise
uncorrelated and cannot be described by a single coherent component.  We
call this `maximally stochastic', where the number of hidden variables
is equal to the number of bins.
The present work uses numerical simulations to quantify stochasticity
above and beyond shot noise statistics.

Following the notation of \citet{seljak04} we define the \textit{bias} between two
populations (e.g., haloes and dark matter), as the ratio of the power spectrum
of the two density fields:
\begin{equation}\label{eqn:bias}
\langle b^{2} (k) \rangle \equiv \frac{P_{h}(k)}{P_{DM}(k)} = \frac{\langle \delta^{2}_{h}(\mathbf{k})\rangle}{\langle \delta^{2}_{DM}(\mathbf{k})\rangle}
\end{equation}
where $\delta_{h}$ and $\delta_{DM}$ are respectively the density fluctuation
of haloes and dark matter and the average is done over the modes.

We also define the \textit{cross-correlation coefficient} $r$, which quantifies
the stochasticity between two density fields. In the case of haloes and dark
matter, it can be expressed as:
\begin{equation}\label{eqn:r}
r(k)=\frac{\langle \delta_{h} (\mathbf{k}) \delta_{DM} (\mathbf{k}) \rangle}{\sqrt{\langle \delta^{2}_{h} (\mathbf{k})\rangle \langle \delta^{2}_{DM} (\mathbf{k}) \rangle}}
\end{equation}
where $-1 \le r \le 1$. If $r=1$ there is no stochasticity, and the
distribution of dark matter can be derived from that of haloes, once the bias
is known.

Bias and cross-correlation coefficient are related through the quantity
$\sigma_{b}$, the relative rms fluctuations in $b$, defined as:
\begin{equation}\label{eqn:sigmab}
\left( \frac{\sigma_{b}}{b} \right)^{2}=\frac{\langle (\delta_{h} - b \delta_{DM} )^{2}\rangle}{\langle\delta^{2}_{h}\rangle}.
\end{equation}
From equations (\ref{eqn:r}) and (\ref{eqn:sigmab}), we have the
\textit{stochastic scatter}:
\begin{equation}\label{eqn:sigmar}
\frac{\sigma_{b}}{b}=\sqrt{2(1-r)}\equiv \mathcal{S}
\end{equation}
Hereafter, when we talk about stochasticity, we will refer to this quantity
$\mathcal{S}$. In fact, even a small departure of $r$ from $1$ implies
not-negligible rms fluctuations.  Stochasticity gives the error when we
estimate the dark matter density
field from the halo distribution only through the bias.

In this work we qualify stochasticity between multiple halo populations and
between haloes and dark matter.

\section{The Simulations}
The simulations were performed using the PMFAST code, a parallel, particle-mesh
code \citep{merz05}. This code was run on the CITA itanium cluster, which has 8
nodes of 4 processors each, and a total of 512GB of RAM, allowing simulations
with many particles and a large dynamic range in mass.  We ran several
simulations using the standard cosmology with $\Omega_{\Lambda}=0.73$, $h=0.7$
and box sizes from $100 h^{-1} \rm{Mpc} $ to $1000  h^{-1}\rm{Mpc}$. The number of particles
ranges from $160^3$ to $1624^3$.  In this paper we use the results obtained
with the largest simulation ($=1000 h^{-1}\rm{Mpc} $  box size and $1624^{3}$
particles). In this simulation the wavemode $k$ can be as small as $k=0.6283
\times 10^{-2} h ~ \rm{Mpc}^{-1}$, so that large scales are well sampled. The
particle mass is $\approx 1.75 \times 10^{10} h^{-1} \rm{M}_{\odot}$.  
This samples haloes both above and below $\rm{M}_{\ast}$, representing
rare and common haloes.
Halo catalogs are defined using the spherical overdensity method
\citep{cole96}: once the density peaks in the particle field are identified,
the overdensity in the cells around the peaks is calculated; haloes are then defined 
to be the spherical regions with overdensity $\delta=178$.
\begin{figure}
\includegraphics[width=\columnwidth]{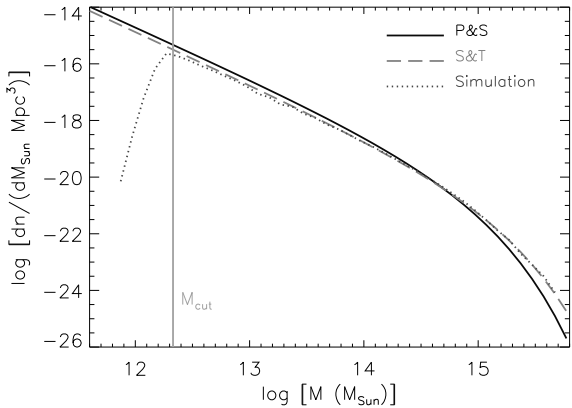}
\caption{Mass function of the halo catalogue (dotted line) compared with the
Press-Shechter (solid black line) and the Sheth \& Tormen (gray long-dashed
line) approximations. Because of the limited resolution in the simulation, we
consider only haloes with a mass higher than $\rm{M}_{cut}=1.5 \times
10^{12} h^{-1}\rm{M}_{\odot}$ (gray vertical solid line). }
\label{fig:PS}
\end{figure}

The mass function of the halo catalogs is then compared both with the
analytical Press \& Schechter \citep[]{press74} and with the Sheth \& Tormen
approximation \citep[]{sheth99}.  Because of the limited resolution in the
simulation, haloes with too few particles have to be excluded from the
catalogue. As shown in Fig. \ref{fig:PS}, small-mass haloes depart from the
Press \& Schechter and the Sheth \& Tormen mass functions. We include in the halo catalogue only haloes above
resolution limits, i.e., above $\rm{M}_{cut}=1.5 \times 10^{12} h^{-1}\rm{M}_{\odot}$.
The final halo catalogue consists of $2.1 \times 10^{6}$ haloes, with masses
ranging from $1.5 \times 10^{12} h^{-1}\rm{M}_{\odot}$ to
$3.7 \times 10^{15} h^{-1}\rm{M}_{\odot} $. 

In order to consider the effect of the shot-noise, we also generated a random
catalogue consisting of as many particles as the number of haloes.  This is the
reference shot-noise catalogue.  It should be noted that the uniformly
distributed random haloes do not obey exclusion: real haloes can not be spaced
closer than a virial radius.  This leads to a slight error in modelling the
shot noise, which we neglect. 

\begin{figure}
\begin{center}
\includegraphics[width=\columnwidth]{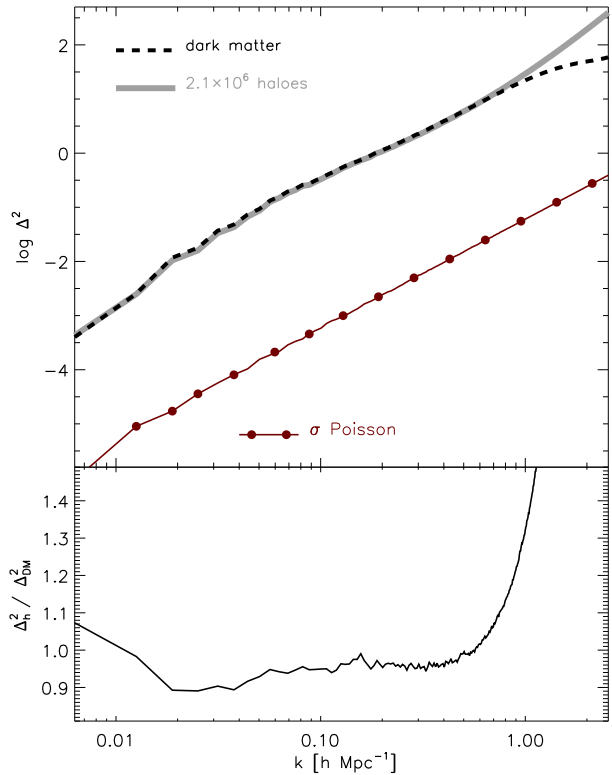}
\caption{\textit{Upper panel}: dimensionless power spectrum of the dark matter (black-dotted line) and
of the halo catalogue (gray-solid line). The Poisson error is shown with the
solid-bullet line. \textit{Lower panel}: ratio of the halo and the dark matter power
spectrum.}  
\label{fig:pktot}
\end{center}
\end{figure}

The dimensionless power spectra of the haloes
and the dark matter are shown in Fig. \ref{fig:pktot} together with the Poisson
counting error. At small scales ($k \gtrsim 1.0 ~ h$ Mpc$^{-1}$) the halo power
spectrum strongly departs from the dark matter power spectrum.  This is because
at those scales  the halo power spectrum is
dominated by the shot noise. As it will be shown in the next section, once we
subtract the shot noise, the halo power spectrum will show the same
small-scales turn-down as the dark matter power spectrum.

\section{Stochasticity between haloes of different mass}

We start by studying the relation between haloes of different masses. The
goal of this section is to use PCA to determine whether different halo
populations are coherent or if stochasticity is present and, if present,
to determine its behaviour. As we will see below, the eigenvalues of
a shot-noise-type field are all the same (`maximally stochastic scenario').  
We want to answer the question: does the stochasticity
between haloes of different mass behave in the same manner? 

\subsection{Applying PCA to the halo covariance matrix}

The first step is to divide our simulated haloes into bins, sorted by mass. We
present the results obtained by dividing the halo catalogue into 6 bins, but we
will show in the next sub-section that the results do not change if we double 
the number of bins.  The bins are chosen with equal number of haloes, such
that the shot noise properties between them are 
similar.  

PCA is applied to the covariance matrix $\sigma_{ij}$
between the six halo bins.  A brief explanation of the way we apply PCA is
given in Appendix \ref{appendixA}.  For each wavemode $k$, the covariance
matrix is given by:
\begin{equation}\label{eqn:covmatrix}
\sigma_{ij}=\langle \delta_{i} \delta_{j} \rangle
\end{equation}
where the indices $i=1,..,n$ and $j=1,..,n$ refer to the halo bins, and where
the average is done over wavemodes during the power spectrum and cross-power spectrum calculation.

In the same way as for the haloes, we divide the random catalogue in bins,
calculate the power spectrum of each bin and the cross-power spectrum between
the bins, and we apply the same procedure of PCA to the covariance matrix from
the random catalogue. 

\begin{figure}
\begin{center}
\includegraphics[width=\columnwidth]{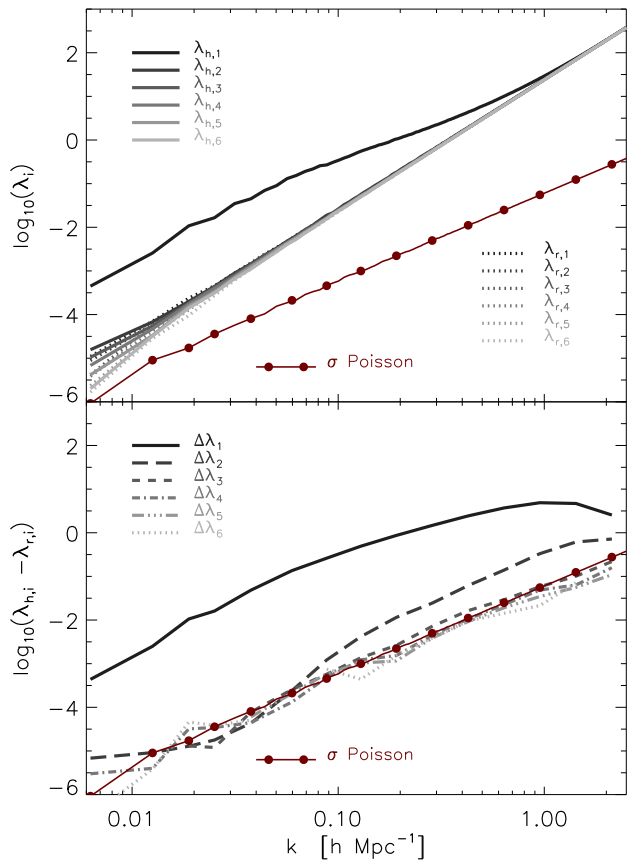}
\caption{\textit{Upper panel}: Eigenvalues of the halo bins covariance matrix (solid thick lines) and of the random bins covariance matrix (dashed thick
Only the first eigenvalue of the halo covariance matrix
  (darkest solid thick line) is significantly higher than the others,
  suggesting the absence of stochasticity in the halo correlation. The Poisson
  error is the same as in the previous figure.  \textit{Lower panel}:
  Difference between the halo bins eigenvalues and the corresponding
  eigenvalues from the random catalogue. The first eigenvalue (solid black
  thick line) is still higher than the others, but we notice here that also the
  second eigenvalue (long-dashed, dark-gray thick line) tends to be
  significantly higher than the others at $k \gtrsim 0.05 ~h$ Mpc$^{-1}$,
  suggesting  that haloes are minimally stochastic.}
\label{fig:eigenvalues_6bins_halos_random}
\end{center}
\end{figure}

In the upper panel of Fig. \ref{fig:eigenvalues_6bins_halos_random} we show the
eigenvalues of the halo covariance matrix (solid lines) compared to the
eigenvalues of the random covariance matrix (dotted lines).  The first
eigenvalue of the halo catalogue is significantly higher than the others, which
in turn are very close to the eigenvalues of the random catalogue. As expected,
the shot-noise eigenvalues are all equal to each other (the scatter at small-k
is due to counting error).

To test if all the eigenvalues but the first one are a consequence of
shot-noise, we subtracted from each halo eigenvalue the corresponding
eigenvalue of the random catalogue. The $\Delta
\lambda_{i}=\lambda_{i,halo}-\lambda_{i,random}$ are plotted in the lower panel
of Fig.  \ref{fig:eigenvalues_6bins_halos_random}. In this case, only $\Delta
\lambda_{1}$ is the dominant one. But we notice that also $\Delta
\lambda_{2}$ is higher than the other differences.  This implies that there is
one additional, and primarily only one, source of stochasticity other than the Poisson
noise. In other words, the detected stochasticity is not a shot-noise like
stochasticity. This result is in agreement with what was found by TB99: 
studying the correlation between four `clans' of galaxies,
TB99 found a \textit{principal component} which traces the matter, which is followed by a second
eigenvalue that is significantly
larger than the remaining two. Their result was obtained for scales around   $k
\sim 0.6 ~ h ~ \rm{Mpc}^{-1} $, which is within the scales where our second
eigenvalue dominates. 

It's clear that this scatter in the relation between different
populations is due to a lack of further information on the populations. Using
PCA we can not further determine on the origin of this scatter, and this is
beyond the goal of the present work. Recent studies have shown that bias between
haloes and dark matter (and therefore between different haloes populations)
could depend on other physical properties other than halo mass, such as halo formation time 
\citep[e.g.][]{gao05, gao07} and concentration \citep[][]{wechsler06}. In principle, one could test any possible
\textit{ingredient} by applying PCA to a $N^{P} \times N^{P}$ covariance matrix,
where $N$ is the number of bins used and $P$ is the number of parameters
included. 

\begin{figure}
\begin{center}
\includegraphics[width=\columnwidth]{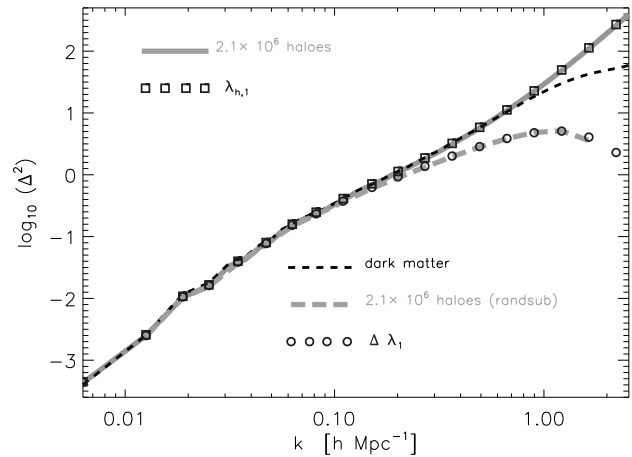}
\caption{$\Delta^{2}$ of the halo catalogue, both before and after random subtraction (respectively gray solid and gray long-dashed lines). The first
  eigenvalue is shown by the symbols, again both before and after random
  subtraction (respectively squares and circles).  The random-subtracted halo
  power spectrum is now lower than the DM power spectrum (thin dashed black line) at
  small scales, because we are neglecting here the 1-halo term.}
\label{fig:delta_eigen}
\end{center}
\end{figure}

Finally, as a check for the fact that most of the signal is contained in the
first eigenvalue, we compare $\lambda_{1}$ with the dimensionless power
spectrum of the complete halo catalogue. As shown in Fig. \ref{fig:delta_eigen}
the first eigenvalue follows the halo $\Delta^{2}$, both in the case of
shot-noise and non shot-noise subtraction. The random-subtracted halo
$\Delta^2$ is now lower than the dark matter $\Delta^2$ at smaller scales, because we
are neglecting here the correlation of structures within haloes (the 1-halo
term).

\subsection{Results with a different number of bins}

\begin{figure}
\begin{center}
\includegraphics[width=\columnwidth]{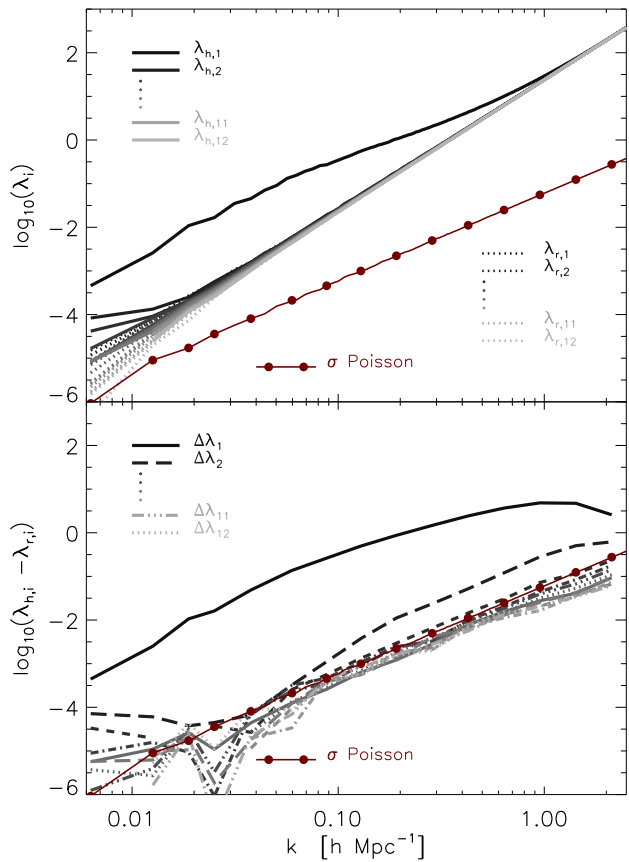}
\caption{Same as Fig. \ref{fig:eigenvalues_6bins_halos_random}, but with a
higher number of halo bins.}
\label{fig:eigenvalues_12bins_halos_random}
\end{center}
\end{figure}

To test how the results described in the previous sub-section depend on the number of bins 
used, we repeated the same exercise using $12$ bins. As show in
Fig.\ref{fig:eigenvalues_12bins_halos_random} the
outcome is the same: we find that there is only one stochastic component other than the shot
noise, i.e., haloes are `minimally stochastic'. The only difference is a
slightly higher scatter in the eigenvalues at small wavemodes, due to
the smaller number of objects in each bin.

\section{Stochasticity between dark matter and halo bins}

In this section we first show how eigenvectors can be used to trace the
bias. We then study the stochasticity between haloes and dark matter, confirming
that stochasticity saturates at scales where the $\Lambda$CDM power spectrum
peaks. Finally, we show how stochasticity can be reduced when the halo density field
is weighted using the \textit{principal component} from the previous section.

\subsection{Bias and Eigenvectors}

\begin{figure}
\begin{center}
\includegraphics[width=\columnwidth]{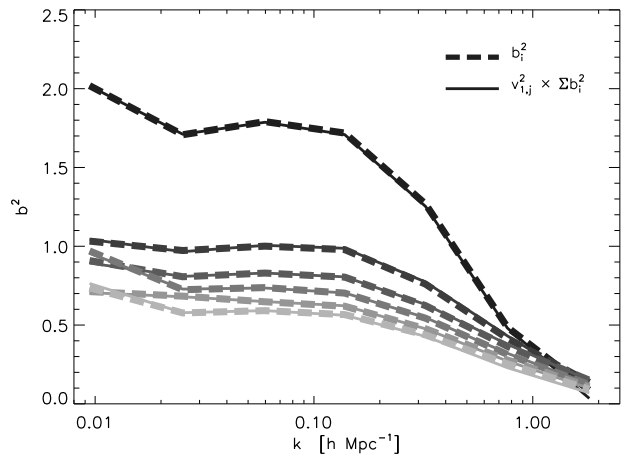}
\caption{Bias between haloes of different mass and dark matter (solid lines),
compared to the components of the first eigenvector. The shot-noise
  has been previously subtracted. The higher biases (darker lines) correspond to bins with
higher mass haloes. } 
\label{fig:bias}
\end{center}
\end{figure}
The bias between the haloes, random subtracted, and the dark matter is shown in
Fig.\ref{fig:bias}, where the highest value of the bias is the one related to
the bin with higher-mass haloes. In the same figure we also plotted the
components of the \textit{principal component}, i.e., the eigenvector corresponding to
the first eigenvalue derived in the previous section. When multiplied by the
sum of all the bias at each $k$, the square of the components of the first
eigenvector follow the bias between the corresponding halo bin and the dark
matter. This is straightforward to demonstrate when $r=1$. In this scenario, in fact, 
the covariance is simply given by:
\begin{equation}\label{eqn:covmatrixbias}
\sigma_{ij}=b_{i}b{j}
\end{equation}
In the two-dimensional case, for example, we have:
\begin{equation}
\sigma_{ij}= \left(
\begin{array}{cc}
b_{1}^{2} & b_{1}b_{2} \\
b_{1}b_{2} & b_{2}^{2} \\
\end{array}
\right)
\end{equation}
and it is straightforward to show that the \textit{principal component} of this matrix
is: $v_{1}=(b_{1}, b_{2})$. The bias between different populations can therefore
be approximated by the eigenvectors of the covariance matrix, at scales
where $r \simeq 1$ (see also TB99).

\subsection{Stochasticity}

We now explore in details the behaviour of stochasticity between haloes and dark matter. 
\begin{figure}
\begin{center}
\includegraphics[width=\columnwidth]{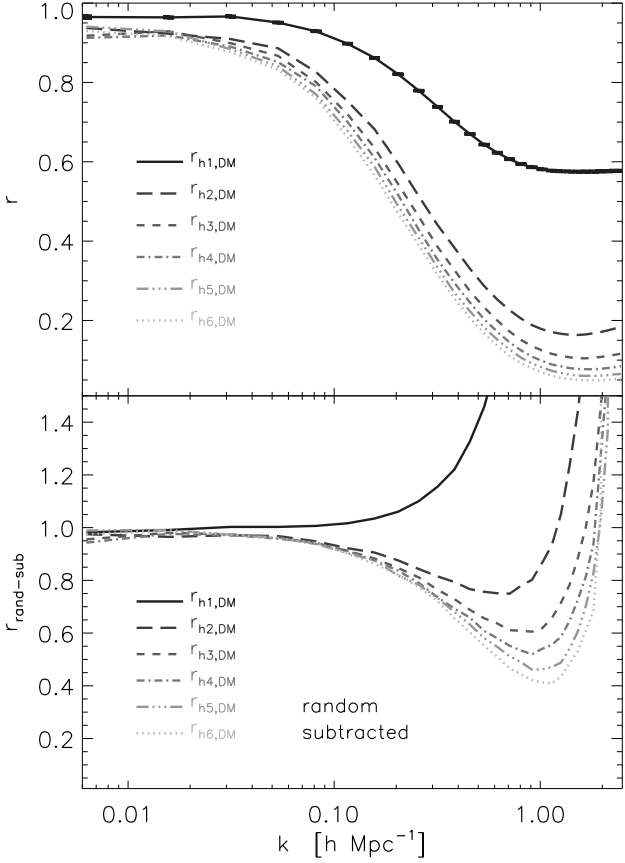}
\caption{Cross-correlation coefficients $r$ between the halo bins and the dark
  matter. In the lower panel $r$ is shot-noise subtracted.} 
\label{fig:rbinsdm}
\end{center}
\end{figure}

\begin{figure}
\begin{center}
\includegraphics[width=\columnwidth]{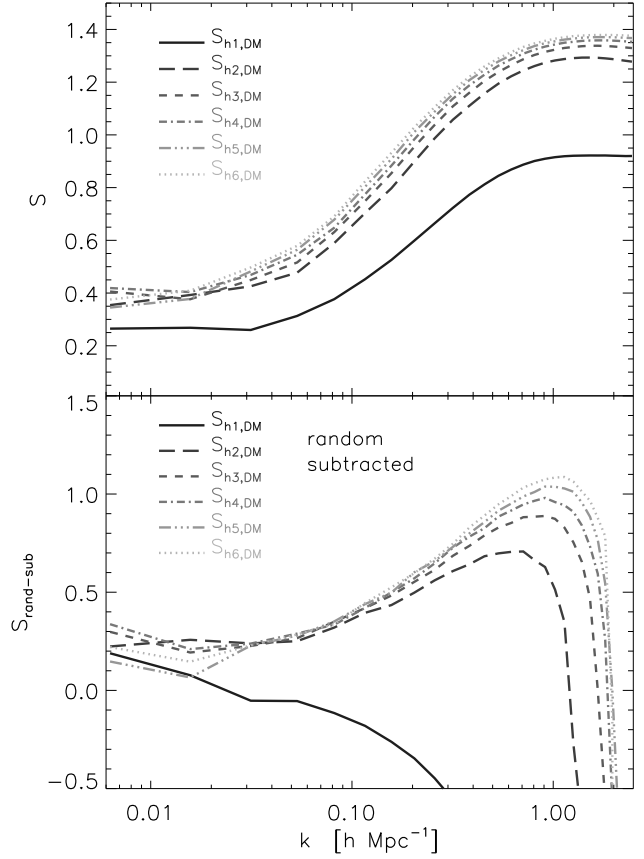}
\caption{$\mathcal{S}=\sqrt{2(1-r)}$ between halo bins and DM. In the lower panel $\mathcal{S}$ is shot-noise subtracted.}
\label{fig:rmsbinsdm}
\end{center}
\end{figure}

The cross-correlation coefficients between each halo bin and the dark matter is
shown in Fig. \ref{fig:rbinsdm} as a function of the wavemode $k$. In all cases,
the
coefficient is close to $1$ at large scales (i.e., lower values of $k$), but it
approaches $0$ at higher values of k, with a more abrupt decrease in the
case of less massive haloes. 

$r$ is indeed close to $1$ at large scales, as it as been always assumed in
works involving the relation between dark matter and haloes (or galaxies), but
io order to verify how good this assumption is, and to determine the effect of
even a small departure of $r$ from unity, we consider the quantity $\mathcal{S}
= \sqrt {2(1-r)} $, which is plotted in Fig. \ref{fig:rmsbinsdm}. Even when $r$
is closest to unity we have $\mathcal{S} \simeq 0.2$, which implies an error
of $\simeq 20 $ per cent in the relation between the halo and the dark matter density
fields. As expected, stochasticity saturates ($\mathcal{S}$ becomes flat) at $k
\la 0.02 ~ h ~ \rm{Mpc}^{-1}$, which corresponds to the peak of the $\Lambda$CDM power spectrum
\citep[e.g.,][]{dodelson96}.  If stochasticity is due to a local process,
one expects its power to be flat at large scales.  The stochasticity thus
decreases with increasing large scale power.  At large scales ($k \la
0.02 ~ h ~ \rm{Mpc}^{-1}$, the power spectrum drops, and one would expect a minimum
in the stochasticity, which we indeed observe.

\begin{figure}
\begin{center}
\includegraphics[width=\columnwidth]{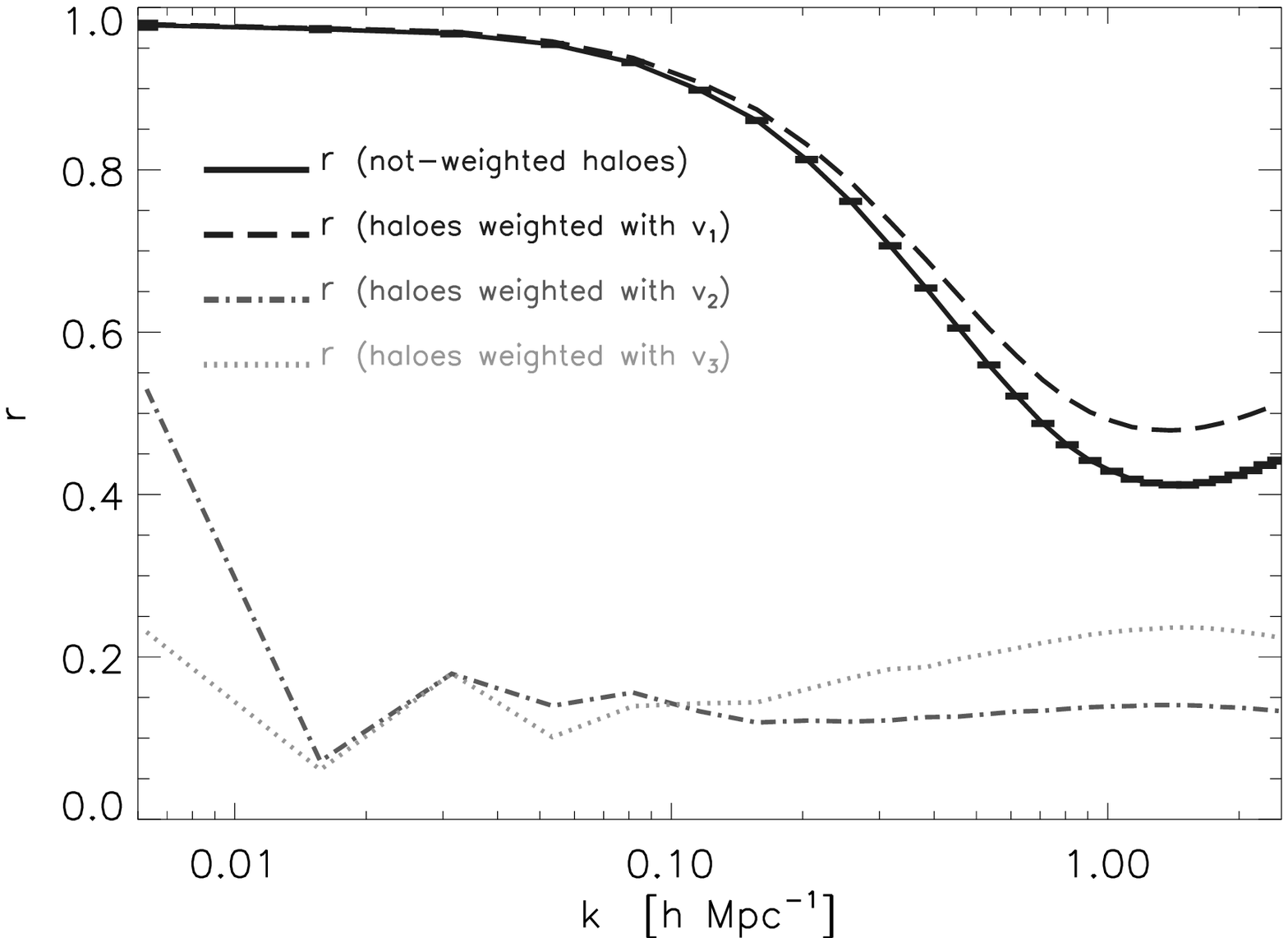}
\includegraphics[width=\columnwidth]{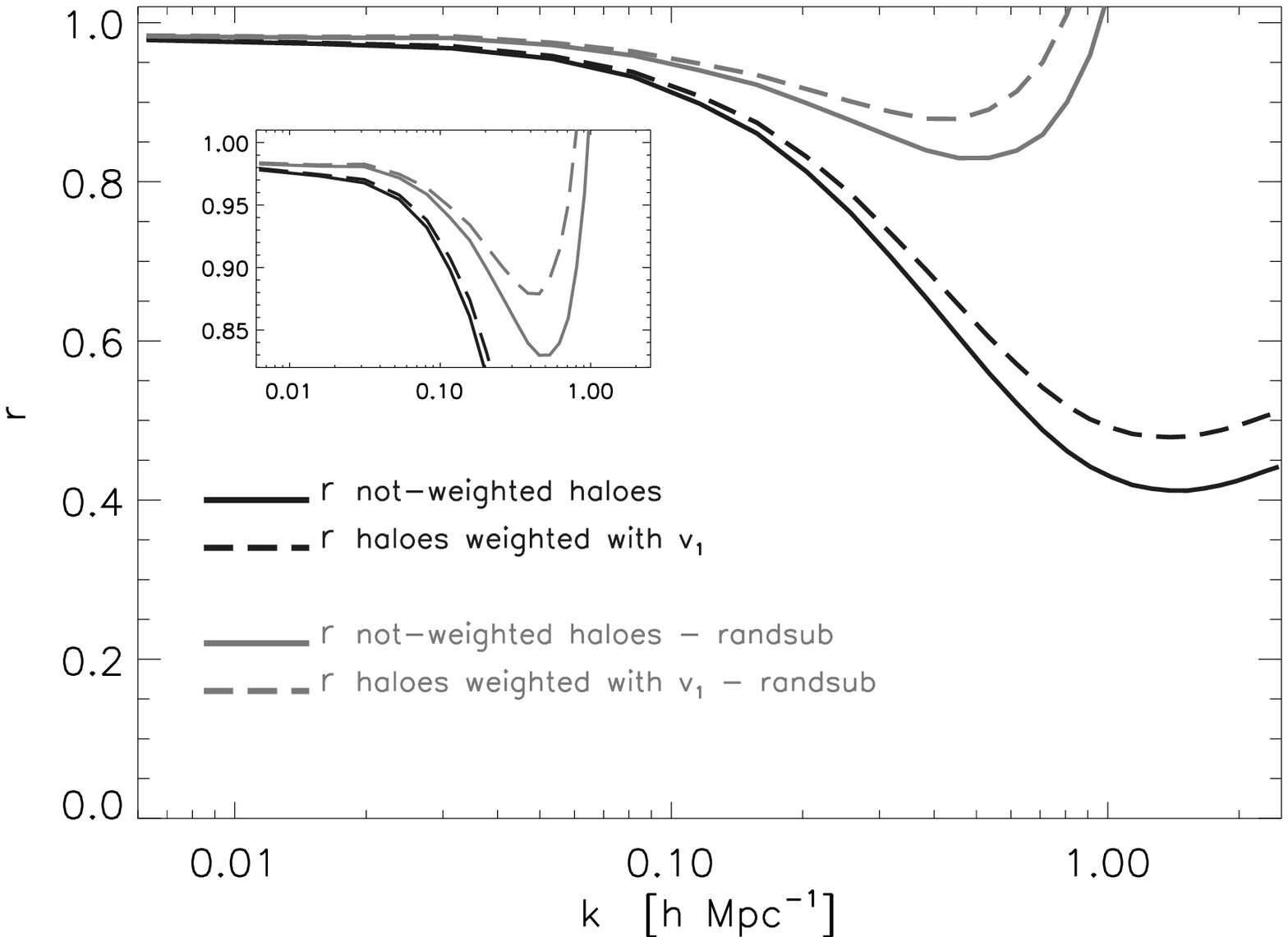}
\caption{Cross-correlation coefficient $r$ between the haloes and the dark
matter, both in the weighted and in the non-weighted case. In the upper panel
the haloes are weighted using the first three eigenvectors, but only the
\textit{principal component} (i.e., the first eigenvector) is significant. In the lower panel the same 
result is shown after the subtraction of the noise.}
\label{fig:rweight}
\end{center}
\end{figure}

\begin{figure}
\begin{center}
\includegraphics[width=\columnwidth]{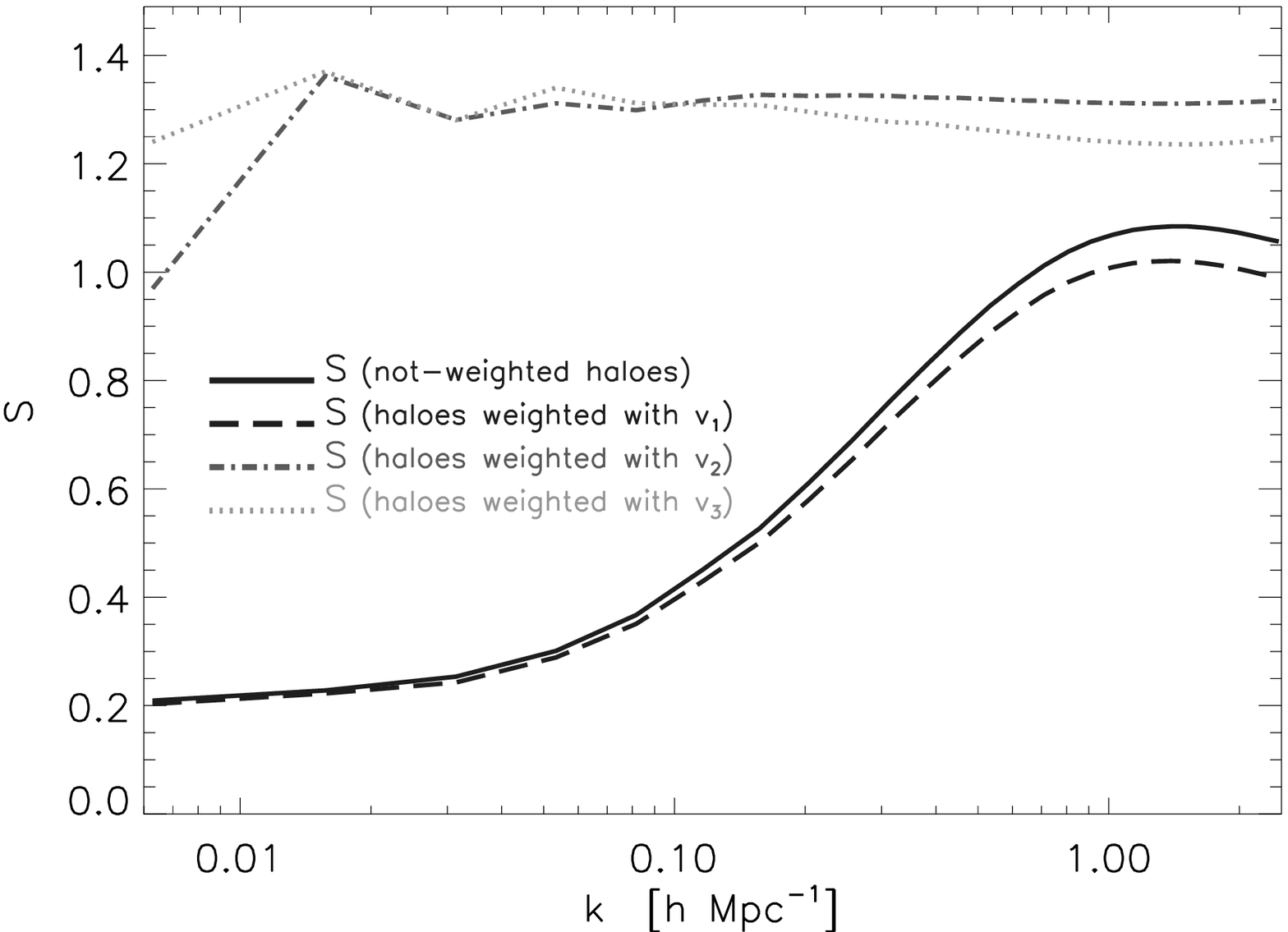}
\includegraphics[width=\columnwidth]{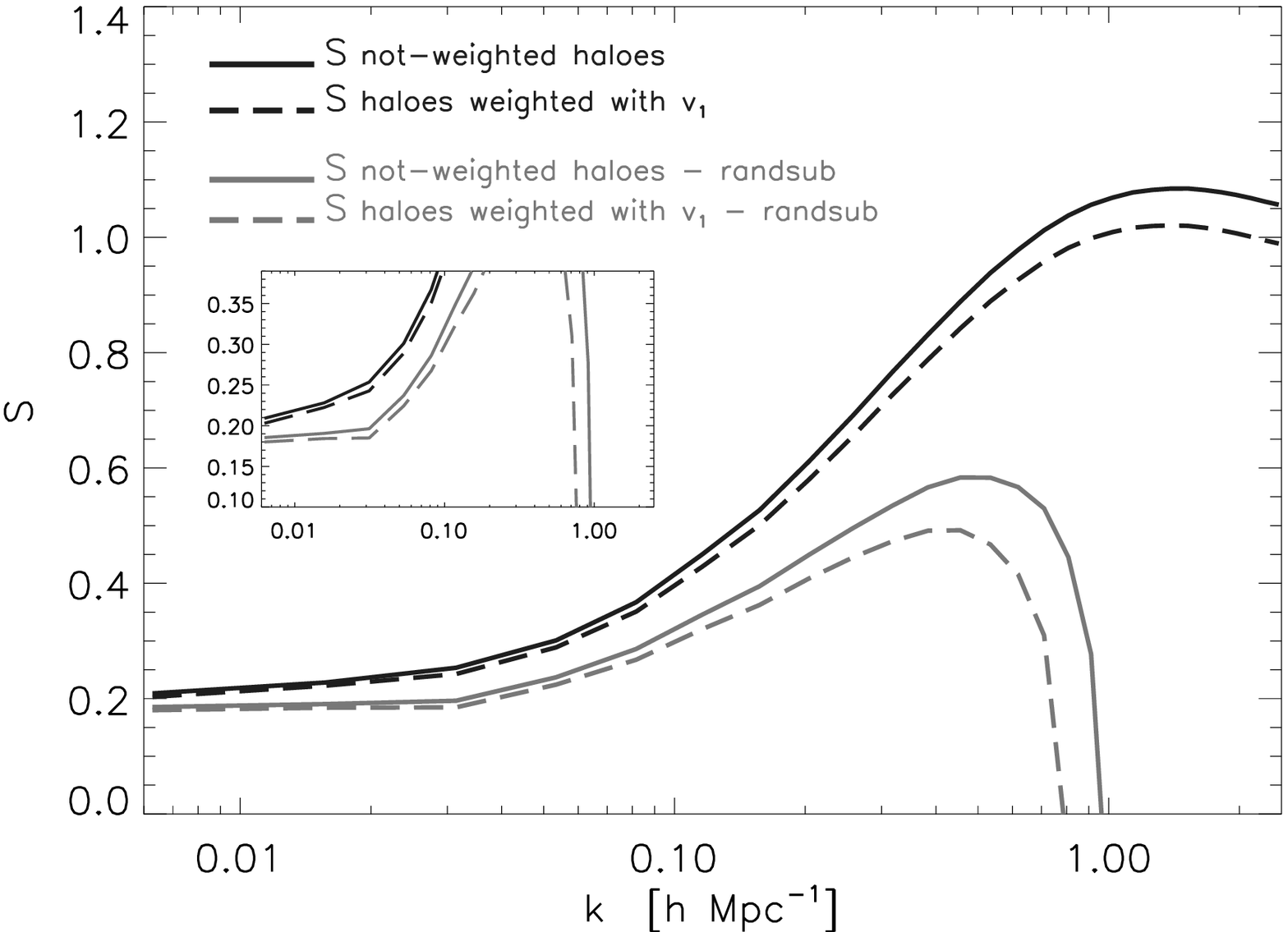}
\caption{Same as previous figure for the quantity $\mathcal{S}=\sqrt{2(1-r)}$. }
\label{fig:rmsweight}
\end{center}
\end{figure}

The values of $r$ and $\mathcal{S}$ are shown both before and after the
shot-noise subtraction.  Notice that, because of the subtraction of the noise,
the value of $r$ can become greater than $1$, in which case $\sqrt{2(1-r)}$ can
be non-real. We still show $\mathcal{S}$, considering that we calculate
$\sqrt{2(1-r)}$ taking
$|1-r|$, and assigning at $\mathcal{S}$ the sign of $(1-r)$. Even after the
shot-noise subtraction, the scatter in the bias saturates at large scales, and
it is of the order of $15$ per cent.

\subsection{Using the \textit{principal component} as weight}

\begin{figure}
\begin{center}
\includegraphics[width=\columnwidth]{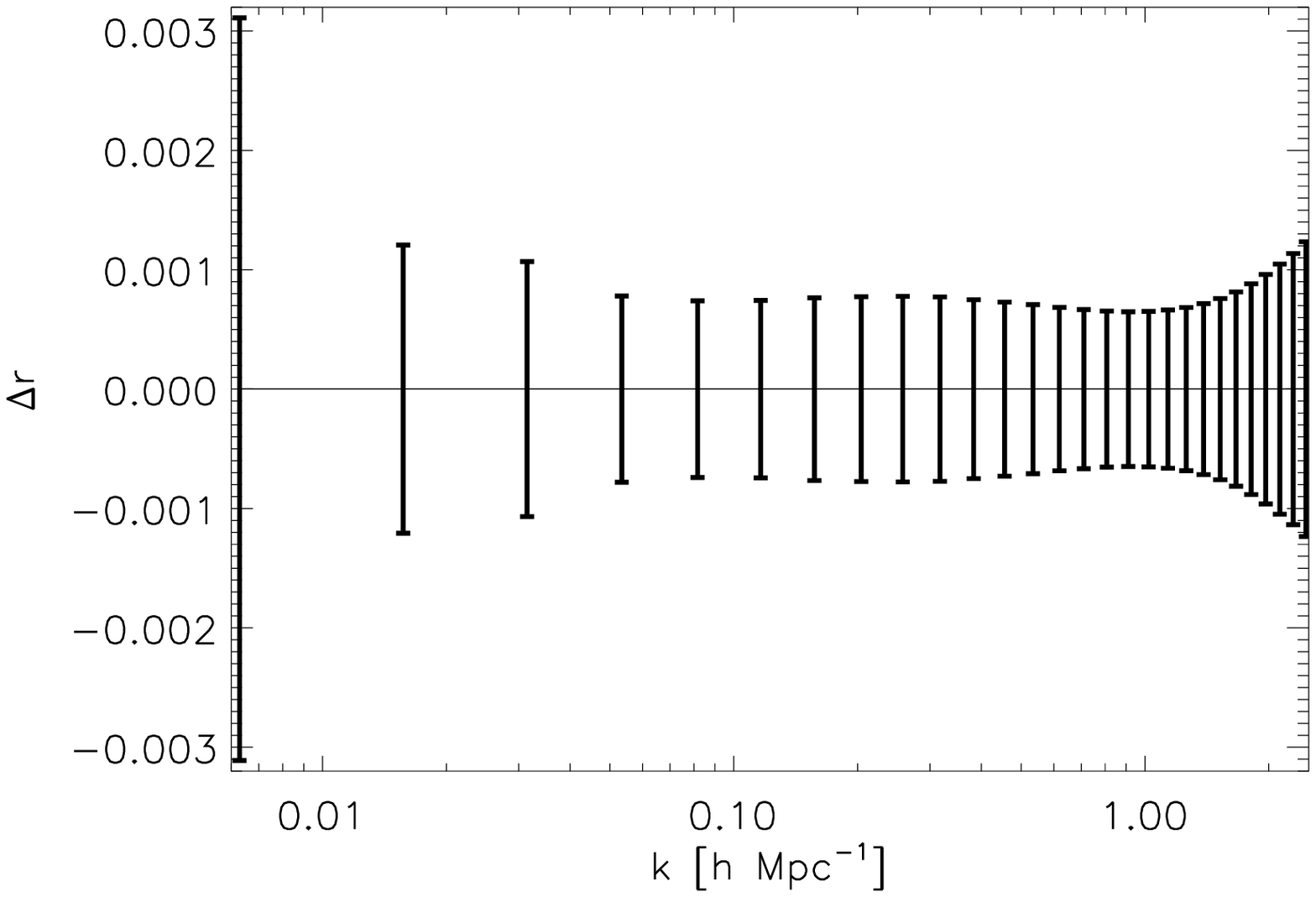}
\caption{Error on the cross-correlation coefficient. For visualization purposes, the error bars have been centered at zero.}
\label{fig:Dr}
\end{center}
\end{figure}

\begin{figure}
\begin{center}
\includegraphics[width=\columnwidth]{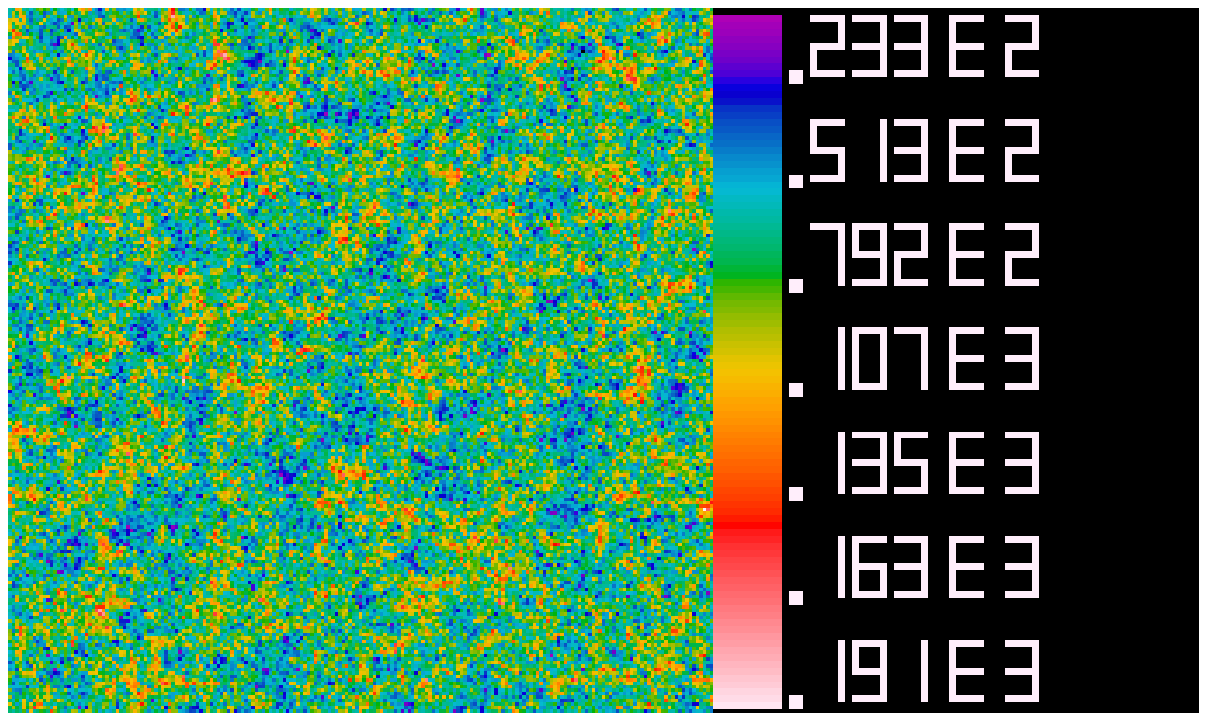}
\includegraphics[width=\columnwidth]{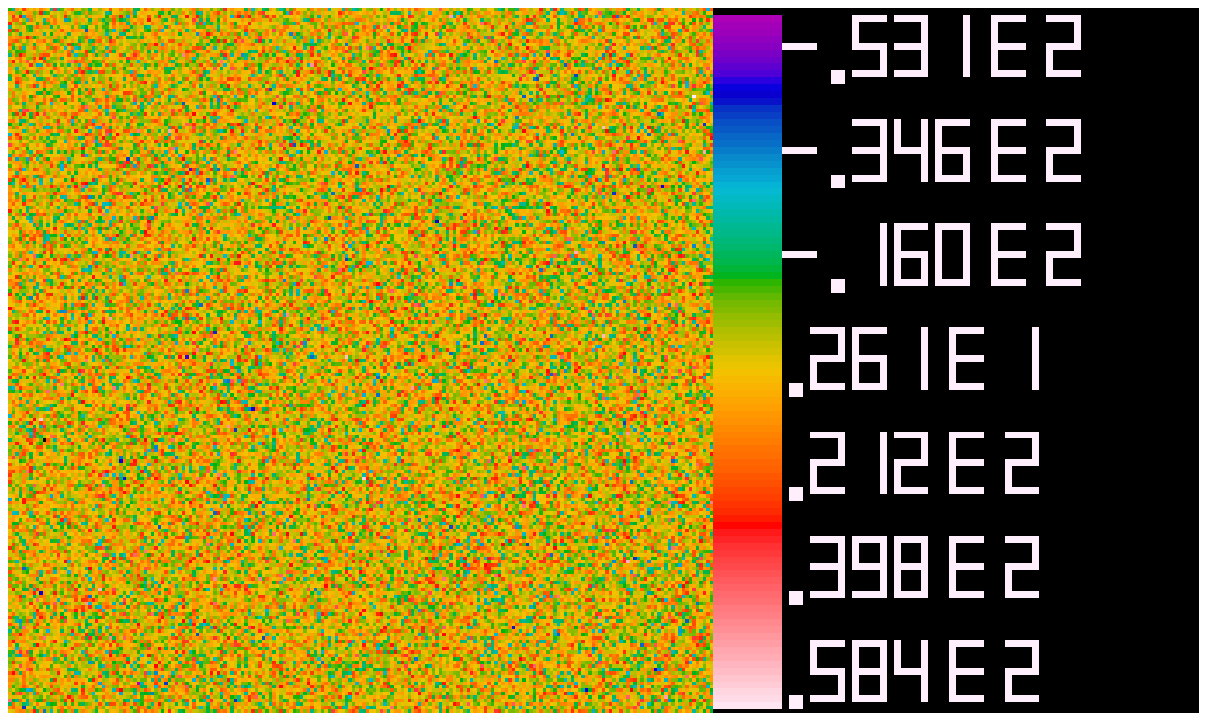}
\includegraphics[width=\columnwidth]{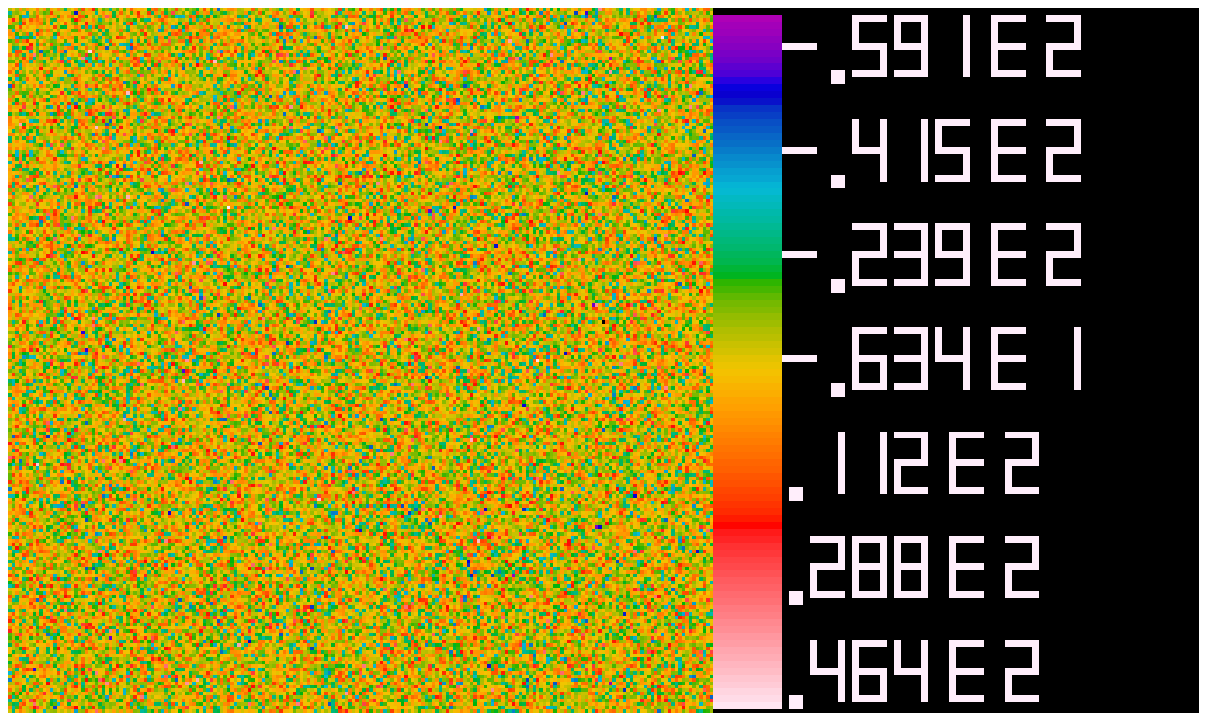}
\caption{Density maps of the haloes. In the first one, the haloes have been weighted using the principal component at $k=0.01885$. The second one is weighted using the second eigenvector and the third maps using the third one. }
\label{fig:maps}
\end{center}
\end{figure}

The results of the previous sub-section are now shown using the `weighted'
values of the halo power spectrum: in the calculation of $\Delta^2$ for the halo catalogue, 
the particle masses are multiplied by a weight depending on which mass-bin they belong to. The
weights are given by the components of the eigenvectors obtained when
diagonalizing the covariance matrix of the $n$ halo bins. 

From statistical theory, assuming to have $k$ independent estimates $e_{i}$ of
the quantity to be measured, each with the associated error $\sigma_{i}$, the
best combined estimate is the weighted mean, given by:
\begin{equation}\label{eqn:defweight}
\overline{e}=\frac{\sum_{i=1}^{k} w_{i} \overline{e_{i}}}{\sum_{i=1}^{k}w_{i}}
\end{equation}
where the weights are given by $w_{i}=1/\sigma_{i}^{2}$. 

In our case the dark matter density field is given by equation (\ref{eqn:bias})
plus an error $\sigma_{noise}$:
\begin{equation}\label{eqn:ourweight}
\delta_{h,i}^2=b_{i}^2  \delta_{DM}^2 \pm \sigma_{noise} \\ 
\end{equation}
where again $i$ refers to the halo bin considered.  Dividing all quantities by
the bias, and using equation (\ref{eqn:defweight}), we obtain:
\begin{equation}\label{eqn:biasweight}
\overline{\delta_{DM}^{2}}=\frac{\sum_{i}^{n}b_{i}^{2}\delta^{2}_{h,i}}{\sum_{i}^{n}b_{i}^{2}}
\end{equation}
We see that in our case the weight for each bin is given by the square of the
corresponding bias.  Using the eigenvectors has advantages over dividing
the power of the bins: when fine bins are used, the shot noise increases.
Unless stochasticity dominates, more information is used when using
the eigenvector, which includes all cross correlations to measure the
bias.

Since the bias is not a measurable quantity in observations, we use as weights
for each bin the corresponding components of the \textit{principal component}. As was
shown at the beginning of this section, $b^2_{i} \equiv v^{2}_{1,j}\times
\sum_{i=1}^{n}b^{2}_{i}$ when $r=1$. Therefore we effectively use as weights
the components of the first eigenvector corresponding to the wavemode
$k=0.01885 ~ h ~ \rm{Mpc}^{-1}$, the wavemode at which the cross-correlation coefficient is closest
to $1$.  At each $k$ there are $6$ eigenvectors, since $r$ is not perfectly
equal to $1$, and each of them has $6$ components. We first weighted the bins
using the components of the \textit{principal component}: the mass of the
particles belonging to the bin $j$ have been multiplied by the $j$th component of
$v_{1,j}$. We repeated this exercise also using the components of second and third
eigenvector.

In Fig. \ref{fig:rweight}, the cross-correlation coefficient between the entire
halo catalogue and the dark matter is shown both in the weighted and
non-weighted case. When weighted using the first eigenvector,  the value of $r$ gets closer to $1$
at every $k$, and the increase is more substantial at larger wavemode. We see
how the weighting done using the eigenvectors corresponding to the secondary
components corresponds to low values of the cross-correlation coefficient, and
this is because these components correspond to noise. 

Again, to see how the weighting process changes stochasticity, we calculate the quantity 
$\sqrt{2(1-r)}$, which is shown in Fig. \ref{fig:rmsweight} as a function of $k$. In this case, the minimum error in
the relation between the entire halo population and the dark matter density field is 
 $\sim 18 $ per cent. 

We need to check if the statistical measurement error in the 
data coming from the
simulation could be the cause of the departure of $r$ from unity. We show that
this can not be the case, and the error on $r$ is small. As derived in appendix
\ref{appendixB}, $\Delta r$ as a function of scale is given by: 
$\Delta r (k)=\langle \epsilon^{2} (\mathbf{k})\rangle / \langle
\delta^{2}_{DM}(\mathbf{k}) \rangle$, where $\epsilon^{2}$ is the random density
field.
$\Delta r$ is shown in Fig. \ref{fig:Dr}, where the error bars have been
centered at $0$. The small values of $\Delta r$ make it insignificant with respect to
$r$, and even when considering possible effects of the error, $r$ can not reach
the unity. 

The density maps of the haloes, properly weighted using the first 3
eigenvectors corresponding to $k=0.01885 ~ h ~ \rm{Mpc}^{-1}$ are shown in Fig. \ref{fig:maps}.

As for the previous section, these last results do not depend on the number of
bins used. To avoid redundancy, we omit the plots.

\section{Summary and Conclusions}

We studied the correlation between
different halo populations and between haloes and dark matter using a PCA
technique. To do so, we used  the output
of a N-body simulation with $1624^{3}$ particles in a $1000 h^{-1}\rm{Mpc} $  box. 
After dividing the halo catalogue into bins sorted by mass, we applied  
PCA to the covariance matrix given by the cross-power spectra of the
halo bins. 

Analyzing the haloes alone, we now understand the stochasticity
between haloes of different mass: we have one dominant principal component, 
and a second component which is small and grows in relative importance
as one approaches the non-linear scales.  This component dominates the
apparent stochasticity. The remaining components are all at the level
expected from random sampling. We call this `minimal stochasticity',
which is the opposite scenario from what one might expect in a shot
noise model, which we called `maximally stochastic'. This result is in agreement
with the conclusions of \citet{tegmark99}, who studied the correlation between different galaxy
populations.

When we consider the relation between
haloes and dark matter, we find that the highest eigenvalue is the best
tracer of dark matter.  Even though this is the best that can be done, the
error is still at least of the order of $15 $ per cent,  even at very large scales, and
at $\lambda  \ga 300 h^{-1} \rm{Mpc} $ stochasticity is saturated as expected: at
this scale the dark matter power spectrum reaches its peak (for a $\Lambda$CDM
cosmology).  
Moreover, we show that the eigenvectors from
PCA are a better estimate for the bias, since it takes into account
all pairs of bins, and has less shot noise.

We also studied the correlation of halo mass bins, and the PCA
eigenvectors, with the total underlying dark matter.  The observation
of galaxies and the measure of their power spectra can provide useful
information on the distribution of dark matter if bias and stochasticity
are properly considered.  When we use the components of the \textit{principal
component} as weights for the calculation of the power spectra of the
haloes, the estimate of the dark matter power spectra improves further.

We argue that this is a better way to properly calculate the galaxy
power spectra and then estimate the underlying dark matter power spectra and to
properly estimate the bias between galaxies of different luminosity and
dark matter. 

\section*{Acknowledgments}
We thank Hugh Merz for the technical support in running the simulations. S.B.
thanks the Department of Astronomy \& Astrophysics at University of Toronto and the Canadian
Institute for Theoretical Astrophysics where most of this work has been carried
out.

\bibliographystyle{mn2e}
\bibliography{bonoli_pen}

\appendix
\section{PCA}\label{appendixA}

Principal Component Analysis is used to identify patterns in data, and to
highlight how data are related to each other. The first step is to calculate
the covariance matrix of the data set. The dimension of the matrix $n$ is equal
to the dimension of the data set.  The covariance
matrix is defined as:
\begin{equation}\label{eqn:defcovmatrix}
\sigma_{ij}\equiv \langle (x_{i} - \mu_{i}) (x_{j} -\mu_{j})  \rangle
\end{equation}
where $i=1,..,n$, $j=1,..,n$ and $\mu$ indicates the mean of the data for the
considered dimension. The main diagonal, $i=j$, is given by the variances of
the data in that dimension, and the matrix is symmetric. 

PCA calculates the eigenvectors and the corresponding
eigenvalues of a covariance matrix. The eigenvector associated with the highest
eigenvalue is the \textit{principal component} of the data set, and it
indicates the main direction along which the data are distributed. The smaller
an eigenvalue, the smaller the `significance' of that component. 

In summary, the number of significant eigenvalues indicates on how many
directions the data are spread. A data set is considered \textit{stochastic} if
more than one eigenvalue are different from zero, and  \textit{deterministic}
if only one eigenvalue is not null, in which case, if you know one data point
you know them all.  In this paper we generalize the nature of stochasticity
to depend on the nature of the eigenvalues.  If all modes have an equal
eigenvalue, as in shot noise, we call it `maximal stochasticity', while
a single mode which dominates the stochasticity is called `minimal stochasticity'.

\section{Errors}\label{appendixB}
\subsection{Spread of the error when binning}\label{appendixB1} 
Throughtout the paper, we often bin along the wavemode $k$. Quantities which
depend on $k$ will then be averaged out within the $k-$bins. If $\sigma_{k,x}$
are the rms fluctuations of the quantity $x(k)$, from basics of statistics the
new error $\sigma_{k_{new},x}$ on each new average $x(k_{new})$, is given by:
\begin{equation}\label{eqn:sigma_binning}
\sigma_{k_{new},x}=\left ( \frac{\sum_{k_{min}}^{k_{max}} \sigma^{2}_{k,x}}{N^{2}_{k,bin}} \right)^{1/2}
\end{equation}
where $k_{min}$ and $k_{max}$ are the extremes of the bin centered in
$k_{new}$, and $N^{2}_{k,bin}$ is the number of wavemodes contained in each
bin.

\subsection{Errors on $r$ and $\mathcal{S}$}\label{appendixB2}
We derive the error on $r$ assuming $r=1$ (No stochasticity null-hypothesis).
The cross-correlation coefficient between haloes and dark matter is given by:
\begin{equation} \label{eqn:r_halodm}
r(k)=\frac{\langle \delta_{h} (\mathbf{k}) \delta_{DM} (\mathbf{k}) \rangle}{\sqrt{\langle \delta^{2}_{h} (\mathbf{k})\rangle \langle \delta^{2}_{DM} (\mathbf{k}) \rangle}}
\end{equation}
If there is no stochasticity, we have $\langle \delta^{2}_{h} \rangle = \langle
\delta^{2}_{DM} \rangle + \langle \epsilon^{2} \rangle $, where $\epsilon^{2}$
is the random density field. The cross-correlation coefficient then becomes:
\begin{equation} \label{eqn:r_halodmrand}
r(k)=\frac{\langle (\delta_{DM} (\mathbf{k}) + \epsilon(\mathbf{k})) \delta_{DM} (\mathbf{k}) \rangle}{\sqrt{\langle \delta^{2}_{DM} (\mathbf{k}) \rangle (\langle \delta^{2}_{h} (\mathbf{k}) \rangle - \langle \epsilon^{2} (\mathbf{k}) \rangle)}}
\end{equation}
where, in the denominator, we subtract the random field from the halo bin.

After some simple arithmetic we get to the final steps which show that, in
the absence of stochasticity, the cross-correlation coefficient is indeed unity:
\begin{equation}\label{eqn:r_unity}
r(k)=\frac{\langle \delta^{2}_{DM} (\mathbf{k})\rangle + \langle \epsilon (\mathbf{k}) \delta_{DM} (\mathbf{k}) \rangle}{\langle \delta^{2}_{DM} (\mathbf{k})\rangle}=\frac{\langle \delta^{2}_{DM} (\mathbf{k})\rangle}{\langle \delta^{2}_{DM} (\mathbf{k})\rangle}=1
\end{equation}
Note that $\langle \epsilon (\mathbf{k}) \delta_{DM} (\mathbf{k}) \rangle=0$,
where here the average is done over various directions of $\mathbf{k}$ within a
simulation.

Now, being $1$ the expectation value of $r$, the error on $r$ is given by:
\begin{eqnarray}\label{eqn:def_dr}
(\Delta r (k))^{2} & \equiv & \langle (r (k)-1)^{2 }\rangle \nonumber \\  
& = & \langle r^{2} (k)\rangle -1  \nonumber \\
& = & \frac{\langle  \delta^{2}_{DM}(\mathbf{k}) (\delta_{DM} (\mathbf{k}) + \epsilon (\mathbf{k}))^{2}  \rangle}{ \langle  \delta^{2}_{DM}(\mathbf{k}) \rangle^{2} } -1 \nonumber \\
& = & \frac{\langle \epsilon^{2} (\mathbf{k})\rangle}{\langle \delta^{2}_{DM}(\mathbf{k}) \rangle}
\end{eqnarray}
where the average is done over an ensamble of simulations.

\label{lastpage}

\end{document}